\documentclass[a4paper,11pt]{article}
\pdfoutput=1 
             
\usepackage{siunitx}
\usepackage{graphicx}
\usepackage{authblk}
\usepackage{amsmath}

\newcommand{\onemicron}{\SI{1.55}{\micro\meter}~}
\newcommand{\tenmicron}{\SI{10.59}{\micro\meter}~}

\graphicspath{{FIGURES/}}

\title{\boldmath Design considerations for an ultrahigh-bandwidth Phase Contrast Imaging system applied to fusion grade devices}

\author[a]{A. Marinoni}
\author[b]{J.C. Rost}
\author[b]{M. Porkolab}
\date{}

\affil[a]{Center for Energy Research, University of California San Diego, San Diego (CA), USA}
\affil[b]{Plasma Science and Fusion Center, Massachusetts Institute of Technology, Cambridge (MA), USA}

\begin{document}
	\maketitle
\flushbottom

\abstract{The PCI diagnostic is an internal reference interferometer that creates an image of absolutely calibrated electron density fluctuations integrated along the line of sight of the probing light beam. While conventional PCI diagnostics installed on fusion experiments worldwide employ light of wavelength equal to \tenmicron, the same system using light at \onemicron wavelength would extend the spectral response in wave-number and frequency by factors of seven and over one hundred, respectively, thereby potentially providing quantitative measurements of the internal structure of density perturbations induced by either turbulent or radio-frequency waves, simultaneously covering ion to electron gyro-radius scales up to the GHz frequency region. Based on a previously developed \onemicron PCI prototype system, constraints to the design for such a diagnostic in fusion grade devices are presented and compared to those faced with the conventional method.}

\section{Introduction and motivation}
\label{sec:intro}

The Phase Contrast Imaging (PCI) method is an optical technique that converts phase shifts induced on the wavefront of probing light travelling through a transparent medium into amplitude variations at the image plane, where they can be imaged by suitable arrays of detectors. Invented in the 1930s by F.~Zernike~\cite{Zernike:Physica1934}, who proved its superior sensitivity compared to competing methods used in cellular microscopy, it was first applied to plasmas in the late 1960s~\cite{Presby:RSI1967} before being brought to experiments worldwide on magnetically confined plasmas~\cite{Weisen:RSI1988,Coda:RSI1992,Lo:RSI1997,Chatterjiee:RSI1997,Mazurenko:PHDthesis,Marinoni:RSI2006,Yu:JINST2017,Tanaka:JJAP1992,Tanaka:RSI2003,Huang:JINST2021}. The high sensitivity of the PCI method is enabled by the fact that it is an interferometer for which both the test and the reference legs travel through the object. In the case of probing light frequency much higher than any plasma frequency and for spatial inhomogeneities larger than the probing light wavelength, the Raman-Nath regime~\cite{Raman-Nath:PIAS:1936} applies and the plasma behaves as a transparent object that induces phase shifts proportional to the probing light wavelength and the electron plasma density integrated along the probing beam line of sight.
%
%
The total phase shift $\phi$ can be decomposed into a contribution given by the stationary plasma density $\phi_0(\mathbf{x})$ and that due to fluctuations $\tilde\phi(\mathbf{x},t)$, which is usually much smaller than the former. The probing light electric field $E_0$ becomes equal to 
\begin{equation}\label{EQ:phiPCI2}
E_{tot}(\mathbf{x},t)= E_0e^{\imath\phi(\mathbf{x},t)}=E_0e^{\imath\phi_0(\mathbf{x})}e^{\imath\tilde\phi(\mathbf{x},t)} \simeq E_{\phi_0}[1+\imath\tilde\phi(\mathbf{x},t)],
\end{equation}  
so that the total electric field, $E_{tot}$, is composed of the reference beam, $E_{\phi_0}$, and a $\pi/2$ phase shifted test beam that has interacted with the fluctuating plasma density, $E_{\phi_0}\imath\tilde{\phi}$. The PCI method employs the fact that the two legs propagate at finite angles with respect to each other and are separated in a suitable focal plane of the imaging system. An additional $\pi/2$ phase shift is then applied by having the reference beam impinge onto an appropriately size \textit{groove}, of power reflectivity $\rho$, dug in an otherwise flat optical component called \textit{phase plate}. The \textit{invisible} phase shifts are thereby made \textit{detectable} as variations in brightness and, to first order in the perturbed phase, can be expressed by relation
\begin{equation}\label{EQ:phiPCI3}
|E(\mathbf{x_{\bot}},t)|^2\simeq \rho|E_{\phi_0}|^2\left[1\pm 2\tilde\phi(\mathbf{x_{\bot}},t)/\sqrt{\rho}\right].
\end{equation}  
Equation \ref{EQ:phiPCI3} shows that the sensitivity of a PCI system is a factor $2/\sqrt{\rho}$ higher than that of a comparable homodyne interferometer operated in optimal detection mode, i.e. with the two legs exactly out of phase, with other parameters being equal. 
\paragraph{Motivation for a near infra-red (NIR) PCI system}\label{SEC:motiv}
While a shorter probing wavelength reduces refraction effects and more easily satisfies the low scattering angle approximation and the Raman-Nath regime, optical aberrations and the quality of optical components have less of an impact on the image at longer probing wavelength. Additionally, the focal spot of a given focusing optic increases at increasing wavelength, which eases manufacturing tolerances on the width of the groove and reduces the power density of the reference beam therein. PCI systems installed on magnetic fusion devices worldwide employ $\rm CO_2$ lasers oscillating on the \SI{10.59}{\micro\meter} wavelength, which is a compromise choice of the requirements above. \\
The probing wavelength plays a major role in determining the wave number and frequency spectrum of plasma fluctuations that the system is able to resolve. Indeed, the fact the light should not be clipped by any of the optics sets the upper limit to the scattering angle at which the test beam propagates with respect to the reference beam. By lowering the probing wavelength from \tenmicron to \onemicron allows one to resolve larger wave numbers, use smaller optics and/or cope with longer beam paths. The frequency bandwidth is given by the rise time of the detectors. A \tenmicron system uses arrays HgCdTe detectors that are cryogenically cooled to minimize noise due to thermally excited current carriers. As a result, the 3dB point is usually located at frequencies less than one or \SIrange{10}{20}{\mega\hertz}, depending on whether the detectors are operated in the photoconductive or photovoltaic regime. The study of faster phenomena therefore requires the use of optical heterodyning techniques requiring dedicated hardware components, careful alignment through the laser intensity modulator as well as prior knowledge of the frequency of the wave to be measured. A \onemicron system uses InGaAs arrays of detectors whose 3dB point is, depending on the size of the element, in the range \SIrange{30}{1000}{\mega\hertz}, thereby increasing the bandwidth by up to three orders of magnitude relative to \tenmicron systems and reducing overhead maintenance costs thanks to room temperature operation. This will be especially advantageous in future large scale devices where recirculating liquid nitrogen circuits would have to be designed and operated due to restricted access to the torus hall. A further advantage offered by InGaAs detector arrays is that their cost is 1\% to 10\% that of a comparable array used in \tenmicron systems, which could be a considerable fraction of the overall budget. Although the signal amplitude linearly decreases with shorter probing light wavelength, the SNR does not degrade thanks to the fact that, when compared to cryogenically cooled HgCdTe detectors, InGaAs detectors feature a two order of magnitude higher normalized detectivity $D^*$ and a much larger saturation power, which significantly lower detector and shot noise. General optical components such as lenses, position sensing detectors, beam profilers, beamsplitter, waveplates or modulators are readily available in the NIR and at 10\% to 30\% the cost of their mid infra-red (MIR) counterparts. Finally, being able to digitize at much higher frequency, \onemicron systems entail larger costs for the acquisition system.          
\section{Design considerations}
\begin{figure}
	\centering
	\includegraphics[width=0.3\textwidth]{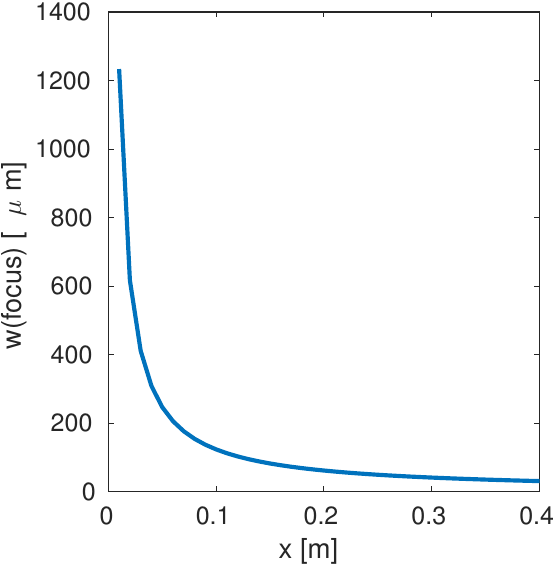}
	\caption{\label{Fig:gaussian_width_focus} Gaussian width at focus for a two-lens system, of 1 m focal length each, that focuses a 
		4 cm wide Gaussian beam of 1.55~$\mu$m wavelength.}
\end{figure}

The design of an imaging system aimed at detecting the spatio-temporal evolution of turbulent structures in fusion devices heavily depends both on the beam path length and on the intensity of the magnetic field that is used to confine the plasma. Based on the ITER plant layout, in future devices we assume a round-trip path \SI{100}{\meter} long. In order to scope the problem at the order-of-magnitude level, we consider confining magnetic fields between \SI{3}{\tesla} and \SI{8}{\tesla}, with the lower value being either the core/edge of a low/mid-field device and the higher value the core of a high field device, along with plasma energies ranging between \SI{0.2}{\kilo\electronvolt} and \SI{10}{\kilo\electronvolt}. Conventional PCI systems employ light beams with a diameter 2$w_0$ of \SIrange{5}{10}{\centi\meter}, so that the minimum wave number resolved by the system is in the vicinity of $\Delta k \simeq \pi/w_0\simeq$\SI{1}{\per\centi\meter}. In present experiments, depending on plasma parameters, this value approaches the bulk of ion Larmor radius scale fluctuations, that are predicted to usually peak around $k_y\rho_D\simeq0.3$. If the diagnostic is to capture the long wavelength limit it should have a resolution $\Delta k\simeq 0.1/\rho_D$ which, for the plasma parameters above, is in the range \SIrange{170}{25}{\per\meter}. Considering the highest wave number limit, true electron scale fluctuations $k_y\rho_e\simeq 1$ would require one to detect extremely fine scale fluctuations which, in the exemplifying case corresponding to the parameters considered above, translate to a high resolution in the range \SIrange{104000}{15000}{\per\meter}. In view of the small Larmor radii in machines at relatively high field, one might use a smaller than usual beam size in the plasma in order to maintain high power on the detectors and small sized optics. This approach would be eased for \onemicron systems, for which the Rayleigh distance for a \SI{1}{\centi\meter} radius beam waist exceeds \SI{200}{\meter}. In a \tenmicron system such a distance would be \SI{30}{\meter} long, i.e. probably shorter than what the beam path length will be in future devices, thereby possibly requiring one to insert intermediate optics along the beam path. 
\paragraph{Refraction from the plasma}
A \tenmicron probing wavelength is short enough to guarantee negligible refraction from the plasma in present experiments. Even in the case of a high density device such as ITER, predictions for the toroidal interferometer-polarimeter diagnostic expect negligible refraction as long as the line averaged electron plasma density is maintained below a few $10^{21}$~\SI{}{\per\meter\cubed}~\cite{VanZeeland:FED2023}. Refraction of a probing light beam travelling entirely in the low field side region of a toroidal device, along any direction, can be treated with reasonable accuracy in spherical geometry. In this case, for a parabolic density profile $n(r) = n(0)[1-(r/a)^2]$ with $a$ being the plasma minor radius, Bouger's law applies and the maximum refraction angle is given by $\theta_{max}=\arcsin{[\omega_{pe}(0)^2/\omega^2]}$~\cite{Shmoys:JAPP1961}. Since, especially in the case of long beam paths, active positioning stabilization systems can compensate only small refraction angles, the equation for $\theta_{max}$ implies that, by reducing the probing wavelength from \tenmicron to \onemicron, the maximum allowable density for negligible refraction increases by almost a factor of fifty to reach $10^{23}$~\SI{}{\per\meter\cubed}, well above levels currently envisioned for disruption mitigation techniques.
\paragraph{Depth of field}\label{SEC:DoF}
The longitudinal resolution at the object plane $\delta z$, or depth of field, is given by imposing that the maximum wave-number resolved by the system $\rm k_{max}$ is such that $\rm k_{max}^2/k_0\delta z \le 1$, a condition that is more easily satisfied at shorter probing wavelength. In the case of ion scale fluctuations, i.e. $k_y\rho_D\simeq 0.1$, and for plasma parameters previously considered, the depth of field at the object plane is in the range \SIrange{0.27}{13}{\kilo\meter} at \onemicron vs \SIrange{40}{2000}{\meter} at \tenmicron. For true electron scale fluctuations, i.e. $k_y\rho_e\simeq 1$, the depth of field reaches values of at least a few centimeters only for the \onemicron system in low to moderate field plasmas and for energies larger than a few keV, i.e. in the core region, while it is at the mm or sub-mm level in all other cases. It is usually advantageous to have the probing beam propagate at small angles with respect to the confining magnetic field lines, in a set-up commonly referred to \textit{tangential launching}. Such a configuration, when used in combination with a properly designed spatial filtering technique, typically allows one to drastically improve the spatial resolution of the system as well as to reduce the integration length thereby making it easier for the depth of field to exceed it~\cite{Marinoni:RSI2006}. In order to roughly estimate relevant quantities, let us consider a beam propagating on the horizontal plane of the device, aimed in such a way that the tangency point with the magnetic surface is located at normalized radius $r/a=3/4$, which is a region where most of the core turbulence is usually localized. In the case no spatial filter is used, the integration length will extend from the tangency point to the plasma edge, or $L_z = 2\sqrt{ (R_0 + a)^2 - (R_0 + 3/4a)^2}$, where $R_0$ is the centroid of the plasma flux surfaces and $a$ the minor radius. Considering the case of $R_0/a=3/2$ and $R_0/a=3$, yields $L_z = 2.2a$ and $L_z = 2.8a$, respectively. The integration length can be reduced by using a spatial filter, for which a study using the geometry of the TCV tokamak showed that a radial resolution $\delta r=0.05a$ is well in hand~\cite{Marinoni:RSI2006}. The integration length then reduces to $L_z = 2\sqrt{ (R_0+3/4a + \delta r)^2 - (R_0 + 3/4a)^2 } = [a,1.2a]$ for spherical and standard aspect ratio tokamaks, respectively. When accounting for an actual equilibrium with magnetic field lines having a finite pitch angle and, optionally, a slant propagation angle of the light beam with respect to the horizontal plane of the machine, the values above can vary by up to 30\%. Given the above, the Raman-Nath regime in the core of a high field device imposes $k_{max}\rho_e$ of the order of 0.03 and 0.1 for the \tenmicron and \onemicron systems, respectively. While a relatively long integration length is beneficial to increase the signal-to-noise ratio, when the depth of field is shorter than the integration length the Raman-Nath regime no longer applies and the Bragg regime is needed to interpret the measurements. However, in case the investigation does not require the system to reveal channel-to-channel phasing information that could be important to untangle the physics at play, it is still possible to infer the power spectrum in defocused systems and it is often done in present day experiments.     
A brief discussion of optical aberrations is warranted. Although working at reduced probing wavelength is beneficial for aberrations because of smaller scattering angles and smaller optics, a given aberration is more critical to the image quality at shorter probing wavelength. Considering the highest wave numbers compatible with the Raman-Nath regime as derived above, we consider a design made of three lenses after the phase plate that are used to provide a transverse magnification equal to 0.1. Including third order Seidel aberrations, at the image plane of the \tenmicron system we estimate a \SI{90}{\micro\meter} wide blur and 3\% as optical path difference normalized to the probing wavelength. For the \onemicron system, despite the fact that the maximum wave number it can resolve is almost three times as large, the scattering angle is about three times smaller, resulting in an estimated \SI{50}{\micro\meter} wide blur and 5\% as optical path difference normalized to the probing wavelength. The impact of surface inaccuracies of various optics was estimated by including ten mirrors, two vacuum interface windows and five lenses. Surface accuracy, expressed in units of \SI{632}{\nano\meter} per inch peak-valley, was specified to 0.1 for all mirrors, 1/4 for NBK-7 lenses as well as for ZnSe lenses and vacuum interface windows, while $\rm BaF_2$ windows were assigned a value of unity. A conservative estimate for the surface of the optical elements illuminated by ray bundles originating from the plasma yields an optical path distance that, normalized to the probing wavelength, is equal to 1.5\% and 8\%, respectively, for the \tenmicron and \onemicron systems.    
\paragraph{Focal spot}\label{SEC:focalspot}
Reducing the wavelength of the probing light beam is disadvantageous to PCI systems because the depth and the width of the phase plate groove must be reduced by the same factor. While the requirements for the depth were met in the prototype \onemicron PCI system~\cite{Marinoni:JINST2022}, those given by the width might be of higher concern. Indeed, while a \SI{2}{\meter} focal length focuses a $2w_0=$~\SI{10}{\centi\meter} wide, $M^2=1.1$, \tenmicron laser beam on a spot size $2w_f=\frac{4M^2\lambda_0 f}{\pi 2w_0}\simeq$~\SI{300}{\micro\meter}, such a value becomes less than \SI{50}{\micro\meter} with an equivalent system using \onemicron wavelength light. This, besides increasing the power density on the phase groove by about a factor of 50, translates into much stricter requirements regarding the uniformity of the phase groove. More specifically, while a focal spot on the phase plate is allowed to move by tens of microns with negligible impact on the signal quality in typical \tenmicron PCI systems, on an equivalent \onemicron system such a perturbation cannot exceed a few micrometers. This value sets the stabilization requirements for the feedback system, as well as on the uniformity of the groove width along its length because the feedback stabilization system will generally allow some level of movement along that direction.
While such a requirement is relaxed by lengthening the focal length of the optics used to focus the beam on the phase plate, increasing this by the ratio of the two probing wavelengths, $10.6/1.55\simeq 7$, will result in the need for a focusing optics large enough to collect the entire beam and with a focal length in the range \SIrange{15}{20}{\meter}, which is expensive if at all possible. A solution involves using relay focusing optics, all having relatively short focal lengths that are commercially available, to create a system having the desired effective focal length. Let us consider a system composed of two focusing optics spaced by a distance $l_1$. The formalism of the Ray Transfer Matrix Analysis~\cite{Gerrard+Burch} can be used to derive the distance $r$ from and the angle of propagation $\theta$ with respect to the optical axis of any ray at distance $l_2$ after the second lens, given the corresponding values before the first focusing optics. In order to simplify the equations we will assume that the two optics have the same focal length $f$, for which  
\[
\begin{bmatrix}
A &	,	& B  \\
C &	,	& D 
\end{bmatrix}
=
\begin{bmatrix}
\frac{f^2+l_1l_2-f(l_1+2l_2)}{f^2} & ,   & l_1+l_2-\frac{l_1l_2}{f}  \\
-\frac{2f-l_1}{f^2}                & ,  & 1-\frac{l_1}{f} 
\end{bmatrix}.
\]
\noindent The $C$ element indicates that the spot-size on the phase plate can be made larger by making $l_1$ slightly longer than $2f$, in such a way as to have an effective focal length equal to $f_{\text{eff}}=f^2/(l_1-2f)$. Not only does this allow a \onemicron system to have wider phase grooves, but deviations of the groove width from the nominal value can be compensated by slightly displacing the two focusing optics. More specifically, the complex beam parameter $q$ transforms as $1/q=(C+D/q_0)/(A+B/q_0)$ and its value on the phase groove can be calculated at focus by using the ABCD matrix given the value $q_0$ at the object plane, located a (large) distance $l_0$ from the first focusing optics. By letting $l_1$ exceed twice the focal length $f$ by a (small) quantity $x$, i.e. $l_1=2f+x$, choosing $l_2$ such that the system is at focus, the ABCD matrix reduces to
\[
\begin{bmatrix}
A &	,	& B  \\
C &	,	& D 
\end{bmatrix}
=
\begin{bmatrix}
0	& ,	& -\frac{f^2}{x}  \\
\frac{x}{f^2}	& ,	& -x\frac{l_0-f}{f^2}-1 
\end{bmatrix}.
\]
\noindent Since the Rayleigh distance for \onemicron systems is large, to very good accuracy we can consider the expanded beam at the object plane as flat, i.e. $1/q_0 = -\imath/z_R$. As a result, the width of the Gaussian beam at focus then becomes equal to
\begin{equation}
w(\text{focus})=\frac{\lambda_0f^2}{\pi w_0 x},
\end{equation}
which shows that, as displayed in figure~\ref{Fig:gaussian_width_focus}, if the displacement parameter $x$ is sufficiently small compared to the focal length $f$, the waist at the focus can be significantly larger than~\SI{50}{\micro\meter}.
%
\paragraph{Feed-back stabilization}\label{SEC:feedback}
%
%
%
Internal reference interferometers are insensitive to mechanical vibrations as long as the trajectory of probing light beam does not deviate from its design path. When optical components are mounted on any vibrating component, the focused beam might move away from the phase groove thereby zeroing the signal. Current PCI systems employ active stabilization system using small galvo mirrors that adjust in real time the position of the beam on the phase plate by making an image of the focal plane on an array of detectors in a quadrant layout configuration. Systems employing light at \onemicron wavelength can benefit from the availability of duo-lateral position sensing detectors (PSD). Indeed, as opposed to quadrant positioning photodiodes that require the beam light to overlap in all quadrants, the performance of lateral effect sensors is independent of beam shape and size and only requires light to impinge on their sensing area. Additionally, while the response of quadrant detectors is linear only for small displacements, that of PSD is linear over the entire range of the sensing area which can be as large as \SI{10}{\milli\meter}. Although PSD in the visible spectral region could be used by \tenmicron systems, they yielded inadequate performance in early tests due to power flicker noise from the HeNe laser used for alignment as well as ambient light disturbance~\cite{Coda:PHD1992}. Both of these noise sources are expected to play a much smaller impact on a \onemicron system using scientific grade Er laser and PSD sensitive only around \onemicron wavelength. 
As for quadrant detectors, the accuracy of the measurements can be negatively affected by the presence of ghost beams due to secondary reflections from other optics. This is due to the fact PSD only measure the centroid of the light impinging on the sensor, which causes large errors when two beamlets are present on the sensor surface. Besides using wedged optics to deviate the trajectory of the ghost beam from that of the main beam and eventually blocking it, this problem can be alleviated by using CCDs that allow to distinguish the main and the ghost beams by measuring signal amplitude of the two light spots, at the expense of complexity and frequency response of the system.
The frequency bandwidth is about 15 kHz, which is comparable to that on existing \tenmicron systems and is adequate for compensating vibrations that, on typical fusion devices, are limited to a few hundred Hertz. In a typical compensation circuit used for PCI systems, such as that installed on the DIII-D tokamak, the dominant uncertainty term in the position of the beam light on the phase groove is given by the uncertainty in the positioning sensor~\cite{Coda:PHD1992}. The spatial resolution of PSD is specified to be less than \SI{100}{\nano\meter}, depending on their size, which has to be convoluted to the spatial power distribution of the light beam. Considering scientific-grade lasers and accounting for an larger than unity magnification between focal plane and PSD, this would meet the design requirements when the focal spot on the phase plate is widened to a few hundreds of microns. Other sources of errors due to the electrical circuit and the scanners are expected to be similar to those in current systems, i.e. negligible.
When compensating for vibration at the focal plane, the image plane undergoes spatial shifts that might compromise the measurements. If we consider a focusing optics at distance $l_0$ from the object plane and a vibrating mirror at distance $D$, either positive or negative, from the object plane, the uncompensated shift at the image plane due to the vibration is equal to $DM\delta\theta$, where $\delta\theta$ is the perturbed direction of propagation of the light beam and $M$ the transverse magnification of the system. Considering a feedback system with galvo mirrors located after the focusing optics and at distance $d$ from the phase plate, then the correction to the propagation in response to $\delta\theta$ is $(F_0/d)\delta\theta$. At the plane corresponding to the position of the galvo mirror, the propagation direction is perturbed by the quantity $[1-(L_0+D)/F_0]\delta\theta$. By applying the correction, the perturbed propagation direction becomes equal to
$\delta\theta_{galvo} = [1-(L_0+D)/F_0 - F_0/d]\delta\theta$.
By propagating till the phase plate followed by a generic ABCD matrix representing the rest of optical systems till the image plane, we obtain that the image is shifted by an amount equal to $\delta r_{image} = M( F_0^2/d + D - F_0 + L_0)\delta\theta$,
as was also derived by Coda~\cite{Coda:PHD1992}.
Considering now a more general case of two mirrors mounted on the vessel, let us orient the reference frame such that the incident light beam propagates along direction $(\hat{x},\hat{y}) = (1,0)$, with the two mirrors oriented as $(\hat{x},\hat{y}) = (-1,1)/\sqrt{2}$ and $(\hat{x},\hat{y}) = (-1,-1)/\sqrt{2}$, respectively, so that, in a vibration free environment, the outgoing light beam propagates along $(\hat{x},\hat{y}) = (-1,0)$ after undergoing two $\pi/2$ reflections on the two mirrors. Let us now consider small vibrations such that the beam undergoes perturbed reflections at angles $\pi/2+\delta\theta_{1,2}$. By writing the two rotation matrices and expanding to second order, the outgoing direction of propagation is given by 
\[
\begin{bmatrix}
\hat{x}_{out}\\ 
\hat{y}_{out}
\end{bmatrix}=
\begin{bmatrix}
\delta\theta_1\delta\theta_2-1+\delta\theta_1^2-\delta\theta_2^2 &	\delta\theta_1+\delta\theta_2 \\
-\delta\theta_1-\delta\theta_2 &		\delta\theta_1\delta\theta_2-1+\delta\theta_1^2-\delta\theta_2^2 
\end{bmatrix}
\begin{bmatrix}
1\\0
\end{bmatrix},
\] 
resulting in a leading order $\delta\theta_{out}=\arctan(\hat{y}/\hat{x})$ that is the first order sum of the individual perturbations. It can be shown that a similar reasoning holds if more vibrating mirrors are added. The root mean squared perturbation to the propagation direction of the beam is therefore much higher when there is significant correlation among the individual perturbations, e.g. when the entire fusion device tilts due to currents induced during a discharge ramp-up phase or a disruption. Considering that in next step devices $L_0 \gg F_0,D,F_0^2/d$, we can approximate $\delta y_{img} = ML_0\delta\theta$. By assuming $L_0=$~\SI{50}{\meter} and overall $\delta\theta =$~\SI{100}{\micro\radian} produced by uncorrelated vibrations of N mirrors, we obtain that the apparent shift in the image is $M\sqrt{N}$ in units of centimeters. This translates in an unacceptable shift at the image plane that would be larger than the element separation along the array of detectors. In the case of a feedback system placed on the launching side of the optics table, such that the galvo mirrors are approximately located at distance $l_0$ before the plasma and vibrations induced at position $l_0-D$, the feedback system would have to compensate for the actual $\delta\theta$ resulting from all the vibrations. The shift at the image plane is equal to $\delta y_{img} = M(D-L_0)\delta\theta$ which, in the case $L_0 - D \gg F_0^2/d$, is quantitatively similar to that for the case of the feedback located on the receiving side of the optics table. The advantage of having a feedback system on the launching side of the optics table, as opposed to the receiving side, is that the distance between the galvo mirror and the phase plate is much longer. This results in the galvo mirrors responding with a larger frequency bandwidth because they have to tilt less to compensate for a given vibration, viz. $\delta\theta$ vs $|1-(L_0+D)/F_0|\delta\theta\gg\delta\theta$.
Given estimates above, we conclude that in future devices with substantial mechanical vibrations, PCI systems relying on vibrating optics will need a second feed-back position stabilizing system at the image plane. This applies to tokamaks when operated in pulsed mode, the requirements for tokamaks in steady-state operation and stellarators will probably be much more forgiving. In the case of a strongly vibrating vessel, we envision a system on the launching side of the optics table to act on the phase plate, while the system compensating the image plane will act on light after is reflected by the phase plate, thereby not interfering with the first system. Given the large beam paths involved, the use of a stabilizing system for the phase plate on the receiving side of the optics table would likely entail stabilizing the laser beam on the galvo mirrors, which would successively stabilize the laser beam on the phase plate. In this case, the first system would have to be limited to a few Hz-bandwidth to compensate shifts due to vibrations during the start-up of a plasma discharge and not interfere with the higher bandwidth system acting on the phase plate. An additional system at the image plane would still be required.
%
%
%
\paragraph{Combined PCI-interferometer}\label{SEC:upgrades}
Using laser light in the near infrared region allows for considerable simplification of the hardware if one were to employ a combined PCI-interferometer. As was demonstrated on DIII-D tokamak~\cite{Davis:RSI2018}, by splitting the light beam coming from the plasma one can build both a PCI and a Mach-Zehnder interferometer using the same port allocation, with the further advantage of having a straightforward absolute calibration of the PCI system when the optical design is such that the interferometer and the PCI responses are made to overlap in wave-number space. In a \tenmicron system, the reference beam of the interferometer should travel the same distance as the plasma beam for phase matching conditions, unless the coherence length of the probing laser is long enough to make the phase noise negligible. In a \onemicron system, the reference leg would instead be sent to a compact fiber bundle, without the need for dedicated optics thus further reducing system complexity and associated costs.     
\section*{Conclusions}
A Phase Contrast Imaging system using \onemicron probing light offers several advantages over conventional \tenmicron systems terms of wave number and frequency bandwidth of the system, cost and availability of individual components as well as maintenance overhead. Technological constraints on the depth of the phase groove can be met, while those on the width can be addressed by dedicated optical design. Preliminary design work shows that aberrations are not expected to be detrimental to the image quality despite the reduced operating wavelength, while surface accuracy of optical components might have a larger impact. Despite a lower signal caused by the shorter probing wavelength, InGaAs detectors are characterized by low-noise and large saturation power densities, resulting in similar SNR to the HgCdTe counterparts. Given the cost and uncertainties of installing a new system on a large fusion device, it would be advisable to first assess the performance of a \onemicron PCI system on either a small plasma machine or alongside a \tenmicron PCI system already installed on a larger scale fusion device, so that the two diagnostics can be quantitatively compared to each other.      

\section*{Acknowledgments}

Work supported in part under US-DoE award DE-SC0018095.



\bibliography{refs}   
\bibliographystyle{JHEP} 

\end{document}